\documentclass[%
reprint,
superscriptaddress,
bibnotes,
amsmath,amssymb,
aps,
prc,
]{revtex4-1}
\usepackage{hyperref}
\usepackage{graphicx}
\usepackage{dcolumn}
\usepackage{bm}

\usepackage{bm, color}

\usepackage{hyperref}
\hypersetup{
	colorlinks=true,
	linkcolor=blue,
	filecolor=magenta,      
	urlcolor=blue,
	citecolor=blue,
}

\begin{document}
	
	\preprint{APS/123-QED}

	\title{Conformal prediction for uncertainties in the neutron star equation of state}
	
	\author{Habib Yousefi Dezdarani}
	\affiliation{Department of Physics, University of Guelph, Guelph, Ontario N1G 2W1, Canada}
	\author{Ryan Curry}
	\affiliation{Department of Physics, University of Guelph, Guelph, Ontario N1G 2W1, Canada}
    \author{Cassandra L. Armstrong}
    \affiliation{Facility for Rare Isotope Beams, Michigan State University, East Lansing, MI, USA.}
    \affiliation{Department of Physics and Astronomy, Michigan State University, East Lansing, MI, USA}
	\author{Alexandros Gezerlis}
	\affiliation{Department of Physics, University of Guelph, Guelph, Ontario N1G 2W1, Canada}

	\date{\today}
	
	\begin{abstract}
 	We study uncertainties in the equation of state of neutron stars using conformal prediction as a distribution-free and model-agnostic method that provides coverage guarantees. In particular, we apply the Conformalized Quantile Regression (CQR) method to posterior samples calculated from Bayesian inference, creating reliable uncertainty bands without assuming a specific form of the underlying distribution. We first construct CQR bands using a polytropic equation of state and the Tolman–Oppenheimer–Volkoff equations. 
    We then apply CQR as a postprocessing step to the posterior distribution of neutron star mass-radius relations provided by the NMMA collaboration and to Quantum Monte Carlo calculations of pure neutron matter. In all cases, empirical coverage studies confirm the robustness of the method.   
	\end{abstract}
	
	\maketitle
	
\section{Introduction}

Neutron stars (NSs) provide a unique laboratory for the study of dense matter. A primary goal in nuclear astrophysics is to determine the equation of state (EOS) of this dense matter. The EOS describes the relationship between pressure and energy density, providing the crucial link between the star's macroscopic properties, such as mass and radius, and the underlying nuclear physics \cite{lattimer_neutron_2001,lattimer_nuclear_2012, ozel_masses_2016, hebeler_equation_2013,watts_colloquium_2016}. At low densities, nuclear interactions can be described by chiral effective field theory (EFT), which provides a systematic expansion of nuclear forces \cite{barrett_ab_2013, carbone_self-consistent_2013, drischler_chiral_2021, hergert_-medium_2016,lynn_quantum_2019,epelbaum_modern_2009,machleidt_chiral_2011,gezerlis_quantum_2013,Gezerlis:2014zia,epelbaum_improved_2015,hammer_nuclear_2020}.  
A common approach to neutron star EOS inference is based on Bayesian analysis, where observational constraints are incorporated through likelihood functions and combined with prior assumptions on the EOS parameters \cite{greif_equation_2019, dietrich_multi-messenger_2020,jiang_psr_2020,al-mamun_combining_2021,raaijmakers_constraints_2021,tang_constraints_2021,jiang_bayesian_2023,lim_neutron_2018}. However,  in Bayesian NS inference, parameter estimation requires repeated solutions of the TOV equations, which makes the inference computationally expensive, particularly when a large number of likelihood evaluations is needed. In recent years, emulators that mimic solutions to the TOV equations have been developed to reduce this computational cost \cite{canizares_accelerated_2015,bonilla_training_2022,liodis_neural-network-based_2024,reed_toward_2024}. In addition, recent work has applied emulators to Auxiliary-Field Diffusion Monte Carlo for nuclear physics \cite{Somasundaram:2024zse,armstrong_emulators_2025, Armstrong:2026aap, Curry:2025pna}.
These theoretical developments are paralleled by developments in multi-messenger astrophysics, where observations provide critical data to inform EOS models. In particular, the NMMA collaboration has combined multi-messenger observations, including gravitational-wave detections of NS mergers, observations of the mass and radius of pulsars by NICER \cite{m_c_miller_et_al_psr_2019}, and heavy pulsar mass measurements with nuclear-theory calculations of the EOS using chiral effective field theory (EFT) predictions at low densities~\cite{dietrich_multi-messenger_2020}. More recently, multi-messenger NS observations have been used to constrain the chiral EFT Hamiltonian's low-energy couplings, highlighting the increasing ability to link microscopic nuclear interactions to astrophysical data \cite{Armstrong:2026aap, somasundaram_inferring_2025}.

With the increasing precision of both nuclear-structure calculations and astrophysical observations, uncertainty quantification (UQ) has become an important challenge in nuclear physics \cite{dobaczewski_error_2014,ireland_enhancing_2015,carlsson_uncertainty_2016,drischler_quantifying_2020, duguet_colloquium_2024,schindler_bayesian_2009,furnstahl_recipe_2015,wesolowski_exploring_2019,furnstahl_quantifying_2015,melendez_bayesian_2017}. Studies have developed Bayesian and Gaussian-process frameworks to propagate order-by-order truncation errors in chiral EFT into the neutron-matter EOS \cite{drischler_how_2020}. In a related effort, microscopic constraints for the EOS and structure of NSs have been explored using a Bayesian model mixing framework \cite{semposki_chiral_2025}. In addition, the NS crust and outer core EOS have been developed with quantified uncertainties, employing a two-dimensional Gaussian process using many-body perturbation theory results based on chiral two- and three-nucleon interactions \cite{Gottling:2025ohe}. 
More recently, a hybrid emulator for in-medium similarity renormalization group calculations of nuclear matter has been developed. This is based on chiral nucleon-nucleon and three-nucleon interactions and an implicit-reduced-basis method emulator built on parametric matrix models, enabling rigorous uncertainty quantification via conformal prediction \cite{Cook:2026yhj}.

Bayesian methods provide a powerful framework for uncertainty quantification, relying on assumptions about prior distributions and underlying models. In this work, we employ conformal prediction (CP) as a postprocessing step applied after Bayesian inference, building on our previous work where CP was introduced and validated for nucleon-nucleon scattering observables \cite{dezdarani_conformal_2026}. CP does not replace the Bayesian approach; rather, it complements it by providing finite-sample and model-agnostic prediction intervals with guaranteed coverage for future observations drawn from the same distribution \cite{Vovk_Gammerman_Shafer_2005,Shafer_Vovk_2008}.
Although CP is a frequentist and model-agnostic framework, it can be combined with Bayesian methods to construct prediction intervals with a user-specified coverage level  (e.g., 95$\%$) without relying on the shape of distributions or model correctness.
The standard CP framework assumes that the data are exchangeable, or equivalently independent and identically distributed (i.i.d.), which is typically a less strict requirement than those imposed by parametric models. A key requirement for the coverage guarantee of CP is data exchangeability, meaning that the joint distribution is invariant under permutations. When this assumption is violated, standard CP coverage may fail. To address this, extensions of CP introduce weighted conformity scores or non-symmetric algorithms that emphasize more recent data points and handle challenges such as distribution drift and asymmetry using tools like weighted quantiles or a residual-based method \cite{Barber_Candes_Ramdas_etal_2023}. Due to its simplicity and robustness, CP has emerged as a reliable approach to uncertainty quantification \cite{Bates_Candes_Lei_etal_2023, Lei_Wasserman_2014,Angelopoulos_Bates_2023, Candes_Lei_Ren_2023, Cauchois_Gupta_Ali_etal_2024,Chernozhukov_Wuthrich_Zhu_2018,Gibbs_Candes_2021,Angelopoulos_Barber_Bates_2025, Klein_Bethune_Ndiaye_etal_2025}. While CP has not been widely employed in physics, recent studies have explored its applications in areas such as gravitational-wave searches \cite{Ashton_Colombo_Harry_etal_2024}, nuclear physics \cite{dezdarani_conformal_2026}, the equation of state of NSs \cite{mendes_certified_2026}, and in the analysis of astrophysical data \cite{Singer_Williams_Ghosh_2025}.

In Sec.~\ref{sec: theory of CP} we review the theory of CP and describe the conformalized quantile regression (CQR) method used throughout this work. In particular, we will explain how quantiles and quantile regression can be used to construct prediction intervals.

In Sec.~\ref{sec: TOY-Baysian}, we first study a toy model based on a Bayesian inference, consisting of a polytropic EOS and the TOV equations. We perform Bayesian inference to calculate posterior distributions of EOS parameters and the corresponding NS observables. Treating CQR as a postprocessing step applied to these posterior samples, we construct 90$\%$ CQR prediction bands and perform model-checking.
Next, we present applications of CQR in realistic settings, where we apply CQR as a postprocessing step to posterior samples of the mass-radius relation of NSs published by the NMMA collaboration \cite{dietrich_multi-messenger_2020}, as well as to posterior samples of the EOS of pure neutron matter obtained from Quantum Monte Carlo calculations published in Ref.~\cite{armstrong_emulators_2025}. For both applications, we perform model-checking analyses to evaluate the empirical coverage of the CQR intervals. 

\section{Conformal Prediction} \label{sec: theory of CP}
Conformal prediction offers a distribution-free approach to quantifying uncertainty. Unlike traditional prediction methods that rely heavily on the assumptions of a specific model, CP builds prediction sets based on the observed behavior of residuals. This model-agnostic property allows CP to be applied across a wide variety of learning algorithms, from simple linear regression to more sophisticated models such as Gaussian Processes \cite{pion_gaussian_2024}. Despite challenges like noise, model misspecification, or limited data, CP ensures that uncertainty is accounted for, providing guarantees on coverage regardless of the predictive model used. For more details, see Ref.~\cite{dezdarani_conformal_2026}, where the methodology and its theoretical foundation were discussed in detail.

Quantile regression plays a central role in constructing CP intervals by allowing us to estimate specific quantiles of a distribution. For a random variable $Y$ with cumulative distribution function $F_Y(y) = P(Y)\leq y$, the $\tau$-th quantile is defined as:
\begin{align}
	Q_Y(\tau)  = F_{Y}^{-1}(\tau) = \text{inf}\{y:F_Y(y)>\tau\}.
\end{align}
This identifies the value below which a proportion $\tau$ of the data lies. For example, the median corresponds to $\tau=0.5$. 

When the distribution of $Y$ depends on another variable $X$, we model conditional quantiles, capturing how quantiles change with $X$. Quantile regression estimates these by expressing the $\tau$-th quantile as a function of $X$. 

Suppose we have data points $(x_i, y_i)$ drawn from an unknown distribution for $i = 1, 2, ... , n$. Instead of just predicting a single value $f(x_{n+1})$ for a new point $x_{n+1}$, the goal is to construct a prediction set $C(x_{n+1})$ such that:
\begin{align}\label{marginal-coverage}
	\mathcal{P}\left[Y_{n+1}\in C(X_{n+1})\right]\geq 1- \alpha,
\end{align}
where $\alpha \in[0,1]$ is a user-specified error level \cite{Angelopoulos_Barber_Bates_2025}. This means that the prediction interval contains the true value $y_{n+1}$ with probability at least $1 - \alpha$ (i.e. $\alpha = 0.05$ gives a 95\% coverage level).  This guarantee is known as the marginal coverage property \cite{Angelopoulos_Barber_Bates_2025}. 
A key advantage of this approach is that it does not assume the predictive model is correct or that the data follows a specific distribution. If the model fits the data poorly, the prediction interval will be wide, reflecting greater uncertainty. In contrast, a better model fit leads to narrower intervals. This adaptability makes CP a practical and reliable tool for uncertainty quantification. 

Although Eq.~(\ref{marginal-coverage}) guarantees coverage on average over all test points, it does not necessarily guarantee coverage at each fixed value of $x$. This stronger property is known as conditional coverage and is defined as
\begin{align}
    \mathcal{P} \left[ Y_{n+1} \in C(X_{n+1}) | X_{n+1}=x \right]\geq1-\alpha.
\end{align}

To construct adaptive and flexible prediction intervals with guaranteed coverage, we employ the split CQR framework. In this approach, the data is divided into two parts: a training set, with $n_{\text{train}}$ data points, used to fit a predictive model $f$, and a calibration set, with $n_{\text {calib}}$ data points, used to calculate conformity scores. The key idea is to first estimate the conditional lower and upper quantiles of $Y$ given $X$, using quantile regression. For a chosen level $\alpha$, these quantiles are:
\begin{align}
	Q_Y(\alpha/2\mid X)\ \text{and}\ Q_Y(1- \alpha/2\mid X).
\end{align}
These quantiles define an initial prediction interval, but to ensure proper coverage, a conformal prediction approach is applied using a calibration set. Specifically, the conformity score is defined as:
\begin{align}
	s_i = \text{max} \{Q_Y(\alpha/2\mid x_i)-y_i,\ y_i - Q_Y(1- \alpha/2\mid x_i) \}, 
\end{align}
which measures how far the value $Y$ lies outside the estimated quantile interval. The next step is to arrange the conformity scores in ascending order, and find $q$ as the $[(1-\alpha)(n_\text{calib} + 1)]$-th element of this ordered sequence. This value $q$ is then used to construct the final CQR prediction set for a new point $x_{n+1}$, which becomes \cite{Angelopoulos_Barber_Bates_2025}:

\begin{align}
	C(x_{n+1}) = 
	[ Q_Y\left(\tfrac{\alpha}{2} \mid x_{n+1} \right) - q,\ 
	Q_Y\left(1 - \tfrac{\alpha}{2} \mid x_{n+1} \right) + q ].
\end{align}
This expands the initial interval as needed to ensure it captures the spread of data with guaranteed coverage, regardless of the underlying distribution of the data. In the applications presented in this work, we use the split CQR procedure to construct prediction bands for the observables of interest. Statistical packages such as the Model Agnostic Prediction Interval Estimator (MAPIE) \cite{noauthor_mapie_nodate} can also be used to calculate CQR intervals.

\section{Applications}\label{sec: TOY-Baysian}
In this section, we demonstrate the application of CQR for uncertainty quantification across three different scenarios. We begin with a toy model to construct and validate CQR prediction bands using the TOV equations. Following this, we apply the method to two realistic datasets as a postprocessing step: mass-radius relations published by the NMMA collaboration in Ref.~\cite{dietrich_multi-messenger_2020} and the EOS of pure neutron matter computed using Quantum Monte Carlo in Ref.~\cite{armstrong_emulators_2025}. In each case, we evaluate the performance of CQR through empirical coverage analysis to show that our results satisfy the marginal coverage property.

\subsection{Toy model: TOV Equations and Bayesian Inference}
In recent years, Bayesian inference has emerged as a powerful tool for studying NS structure, enabling the combination of NS models with constraints from astrophysical observations \cite{greif_equation_2019, dietrich_multi-messenger_2020,jiang_psr_2020,al-mamun_combining_2021,raaijmakers_constraints_2021,tang_constraints_2021,jiang_bayesian_2023,lim_neutron_2018}.
Here, we present a toy model for NS structure, based on a Bayesian inference framework, designed to illustrate how CQR can be applied to Bayesian EOS models.
We employ a polytropic EOS and perform Bayesian inference on its parameters. The CQR method is subsequently applied as a postprocessing step to the posterior samples in order to construct prediction intervals with guaranteed coverage. Finally, we assess the performance of this framework by computing empirical coverage across a range of $1-\alpha$ and by examining the stability of the empirical coverage distributions. The purpose of this setup is not to provide a realistic interpretation of NS structure, but rather to present an example demonstrating how CQR can be integrated with Bayesian inference for uncertainty quantification in EOSs.

To model the structure of NSs, we begin with a polytropic EOS
\begin{align}
P(\rho) = K\rho^{\Gamma}, 
\end{align}
where $\rho$ is the mass density, $P$ is the pressure, $\Gamma$ is defined as the adiabatic constant, and $K$ is a constant. Using the first law of thermodynamics, one can calculate the energy density as,
\begin{align}
    \epsilon(\rho) = \rho c^2 \ + \ \frac{P}{\Gamma - 1} .
\end{align}
In general relativity, the structure of a static, spherically symmetric NS can be determined by combining the
Einstein field equations with an equation describing hydrostatic equilibrium, which gives the TOV equations \cite{alexandros_gezerlis_numerical_nodate},

\begin{align}
\frac{dP(r)}{dr} &=
-\frac{G\, m(r)\, \epsilon(r)}{c^2r^2}
\left[1 + \frac{P(r)}{\epsilon(r)}\right] \\
&\quad\times
\left[1 + \frac{4\pi r^3 P(r)}{m(r)c^2}\right]  \nonumber
\left[1 - \frac{2G m(r)}{c^2 r}\right]^{-1} \\
\frac{dm(r)}{dr} &= \frac{4\pi r^2 \epsilon(r)}{c^2}.
\end{align}
Here $r$ denotes the radial coordinate measured from the center of the star, $G$ is Newton's gravitational constant, and $c$ is the speed of light.

\begin{figure}[b]
    \includegraphics[width=0.5\textwidth]{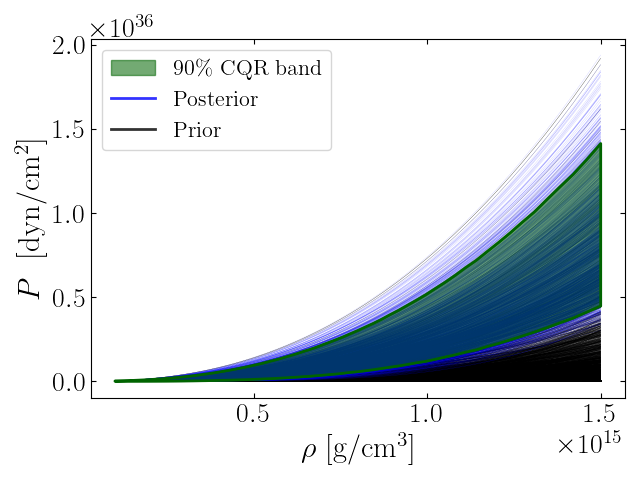}
    \caption{Posterior and prior distributions of polytropic EOS curves with the corresponding CQR bands showing pressure as a function of mass density. Black curves represent $10^4$ samples drawn from the prior distribution, and blue curves represent the same EOS samples drawn from the posterior distribution, calculated after taking three heavy pulsar measurements. CQR is then applied as a postprocessing step to the posterior samples. The resulting $90\%$ CQR prediction intervals are shown as the green shaded region. }   
    \label{fig: toy-cqr-EOS}    
\end{figure}

Our goal is to infer the posterior distribution of the EOS parameters constructed from prior information and a likelihood based on mass observations. In our Bayesian setup, the EOS parameters $\theta = (\ln K, \Gamma)$ are treated as random variables. We assign independent uniform priors to these parameters, chosen to be broad and weakly informative, as shown in Fig.~\ref{fig: toy-cqr-EOS}. Specifically, the adiabatic index is 
\begin{align}
    \Gamma \approx \mathcal{U} (2, 5),
\end{align}
 and for the polytropic constant, we place a uniform prior on its logarithm,
\begin{align}
    \ln K \approx \mathcal{U}(-40, 5).
\end{align}

\begin{figure}[t]
    \centering
    \includegraphics[width=0.5\textwidth]{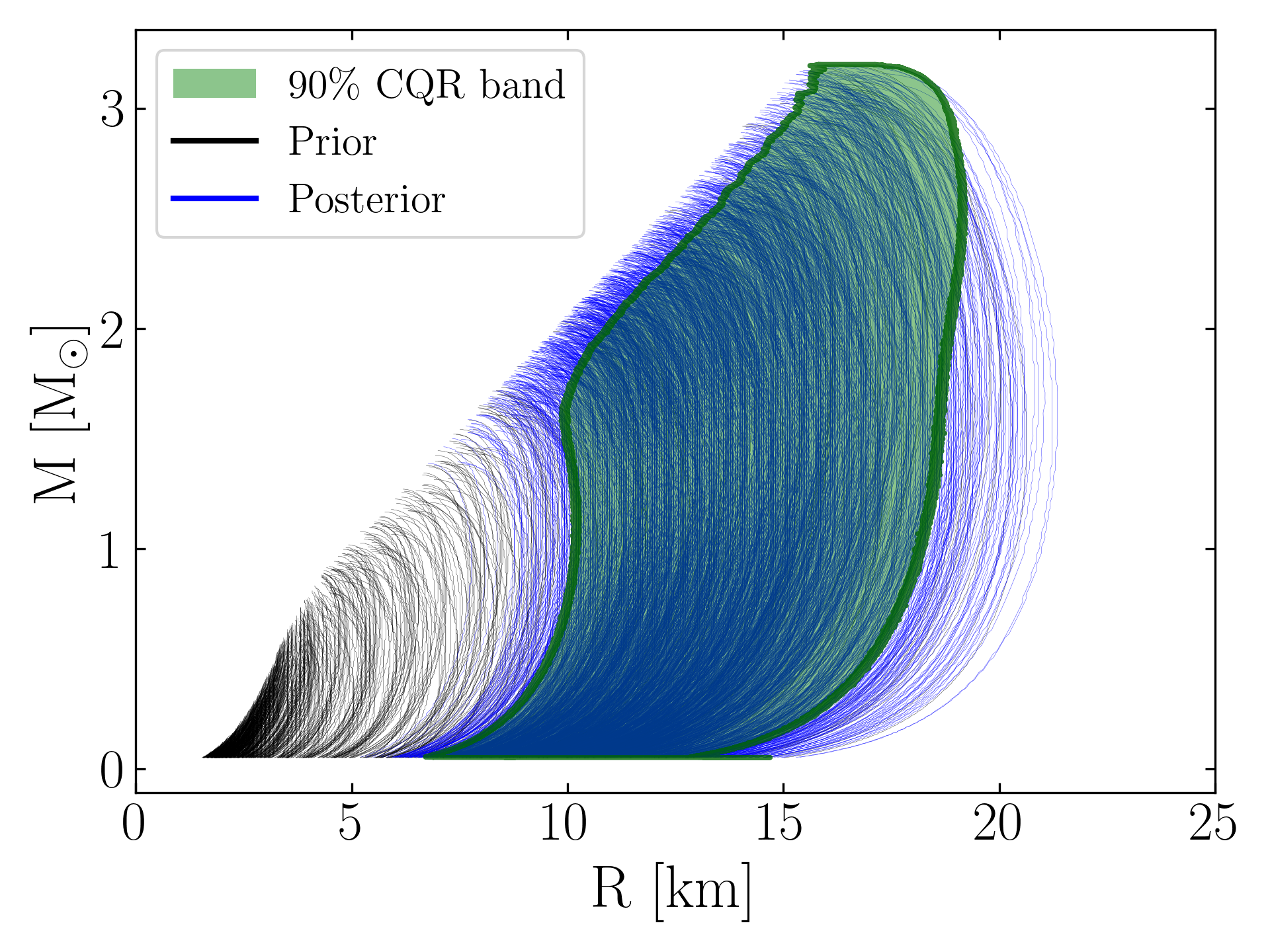}
    \caption{Uncertainty bands for mass-radius relations computed using the polytropic EOS. The black curves represent the prior predictive distribution sampled from broad uniform priors on the EOS parameters ($\Gamma$ and ln$\ K$). The blue curves correspond to those calculated from posterior samples conditioned on three heavy-pulsar mass measurements. The green shaded region illustrates the $90\%$ CQR band; the darker green boundary is included only to guide the eye.}
    \label{fig: toy-cqr-mr}
\end{figure}

We chose these prior ranges to be physically agnostic, covering the broad variations observed in Ref.~\cite{read_constraints_2009}. We also impose the causality condition on EOSs in our prior distribution,
\begin{align}\label{eq: causality}
 \frac{dp}{d\epsilon}=\frac{c_s^2}{c^2}  \leq 1,
\end{align}
where $c_s$ is the speed of sound. We impose these physical constraints at the prior stage. For each parameter set $\theta = [\ln K,\Gamma]$ drawn from the prior distribution, we solve the TOV equations over a broad range of central densities and construct the corresponding mass-radius sequence. We increased the range of central densities and checked that further increasing it does not affect the maximum-mass configurations. We identify the maximum-mass configuration and retain only the stable branch up to this turning point. EOSs that violate causality are rejected. We additionally impose the theoretical upper bound $M_{\text{max}} \leq 3.2~M_{\odot}$ \cite{rhoades_maximum_1974} to remove EOSs that predict unrealistically large neutron-star maximum masses.

This weakly informative prior ensures that the resulting posteriors are primarily informed by the likelihood rather than being merely a result of prior modeling bias.
In Fig.~\ref{fig: toy-cqr-EOS} we show EOS curves, drawn from the prior distribution, for pressure as a function of mass-density. Because the priors on ln$\ K$ and $\Gamma$ are chosen to be weakly informative, the resulting EOS curves cover a wide range of pressure values for a given mass density.

\begin{table}[t]
\centering
\begin{tabular}{lcc}
Object & $M^{\mathrm{obs}}\,[M_\odot]$ & Reference \\
\hline 
Cyg X-2 & $1.71_{-0.21}^{+0.21}$  & \cite{casares_mass_2010}\vspace{0.08cm} \\
4U 1822$-$371 & $1.96_{-0.36}^{+0.35}$  & \cite{munoz-darias_k-correction_2005}\vspace{0.08cm} \\
OAO 1657$-$415 & $1.42_{-0.26}^{+0.26}$ & \cite{mason_evolution_2012} \\
\hline
\end{tabular}
\caption{Neutron star mass measurements and $1\sigma$ uncertainties taken from Ref.~\cite{lattimer_nuclear_2012}.}
\label{tab:pulsar_masses}
\end{table}
\begin{figure}[!b]
    \centering
    \includegraphics[width=0.5\textwidth]{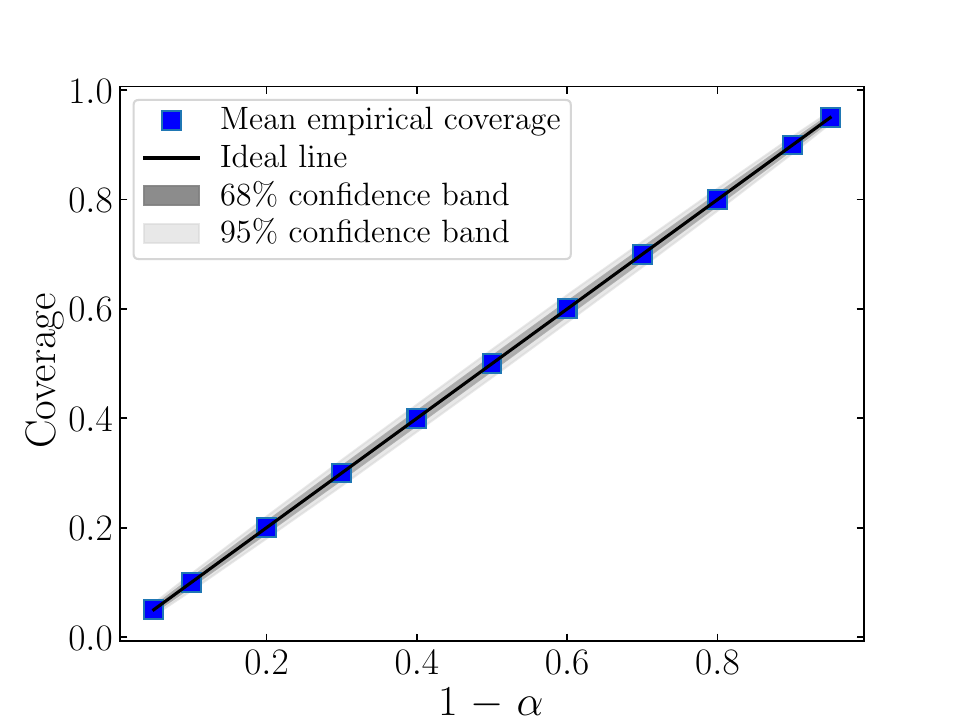}
    \caption{Empirical coverage of CQR prediction intervals for the polytropic EOSs at a fixed mass-density of $2.0\ \times 10^{14} \text{g/cm$^3$}$ as a function of $1-\alpha$. For each value of $\alpha$, the blue squares indicate the mean empirical coverage over $n_{\text{split}}=1000$. The solid black line represents the ideal coverage line. The light and dark shaded regions correspond to the $68\% \ (1\sigma)$ and $95\% \ (1.96\ \sigma)$ confidence bands calculated from the distribution of empirical coverage across splits. The close agreement with the ideal line demonstrates the robustness of the CQR bands.}    
    \label{fig: coverage}
\end{figure}

To update our prior beliefs using observational data, we incorporate the astrophysical dataset $\mathcal{D}$, consisting of three neutron star mass measurements, which is shown in Table \ref{tab:pulsar_masses}. These measurements are chosen arbitrarily for this analysis and were not selected according to a specific observational criterion. We then construct the likelihood following Ref.~\cite{jiang_bayesian_2023} as,  
\begin{align}
    \mathcal{L}(\mathcal{D}\mid \theta) = \prod_{i=1}^3 \frac{1}{2} \left[
    1 \ + \text{erf} \left(\frac{ M_{\mathrm{TOV}}(\theta) - \bar{M}_i}{\sqrt{2}\ \sigma_i}\right)
    \right],
\end{align}
where
\begin{align}
\text{erf}(x):=\frac{2}{\sqrt{\pi}}\int_0^x \ e^{-t^2} \ dt, 
\end{align}
while $\bar{M}_i$ and $\sigma_i$ are the mean and standard deviation of the $i$-th mass measurement, respectively. For a given set of EOS parameters $\theta$, we compute the theoretical maximum NS mass, $M_{\text{TOV}}(\theta)$.
Then, applying Bayes' theorem, the posterior distribution of the EOS parameters is given by
\begin{align}
    \mathcal{P}(\theta|\mathcal{D})  \propto \mathcal{L(\mathcal{D|\theta})}\ \mathcal{P}(\theta),
\end{align}
where $\mathcal{P}(\theta)$ denotes the prior distribution. 

After incorporating observational data through the likelihood, we calculate the posterior distribution $\mathcal{P}(\theta|\mathcal{D})$. We then draw $10^4$ samples from this posterior distribution $\mathcal{P}(\theta|\mathcal{D})$ and use them to generate posterior EOS curves.

In Fig.~\ref{fig: toy-cqr-EOS}, we show the resulting pressure as a function of mass density for the polytropic EOS. 
Once the posterior EOS samples are calculated, we apply CQR as a postprocessing step to quantify uncertainty. At each fixed mass density value $\rho$, the posterior EOS samples are used to form prediction intervals for the pressure $P(\rho)$. This procedure is then repeated across the full density range to construct the $90\%$ CQR band for the EOS.

\begin{figure}[t]
    \includegraphics[width=0.48\textwidth]{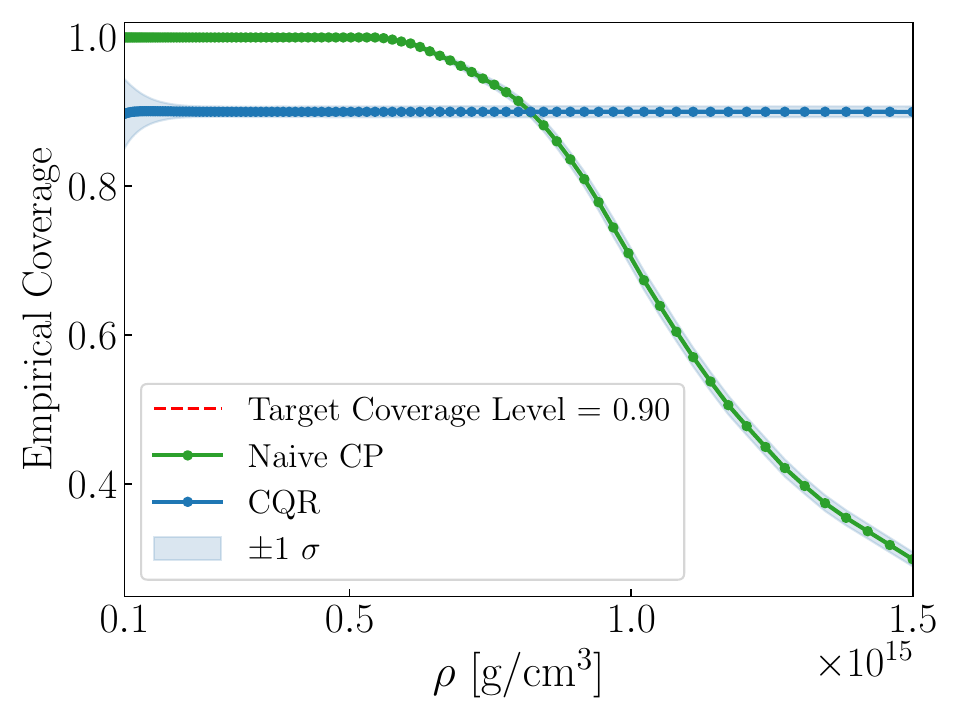}
    \caption{Conditional empirical coverage of naive CP and CQR prediction intervals for the polytropic EOS as a function of mass density. Here we use $10^4$ posterior samples. The green and blue points represent the mean empirical coverage for naive CP and CQR, respectively, while the shaded regions illustrate the corresponding uncertainty bands. The red dashed line indicates the target coverage level ($1-\alpha=0.90$). Naive CP overcovers at low densities and undercovers at high densities, whereas CQR remains close to the target coverage across the full density range by adapting to the changing spread of the posterior samples.}   
    \label{fig: conditional coverage}
\end{figure}
\begin{figure}[b]
    \centering
    \includegraphics[width=0.48\textwidth]{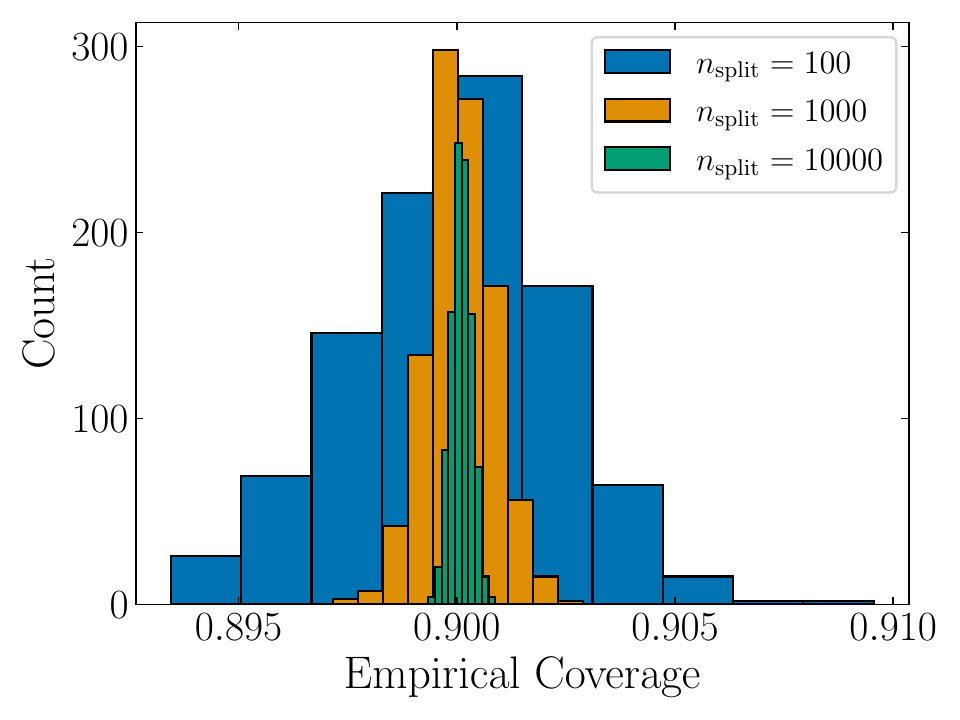}
    \caption{Empirical coverage distribution of the CQR prediction bands for the polytropic EOS at a fixed $\rho = 2.0\ \times10^{14}g/cm^3$. A dataset of $2000$ samples is randomly split into training, calibration, and validation ($n_{\text{train}}=900$, $n_{\text{calib}}=900$, and $n_{\text{valid}}=200$). For each choice of $n_{\text{split}}=100, 1000, 10000$, the procedure is repeated over $n_{\text{trial}}=1000$ independent trials, resulting in a distribution of empirical coverage values around the target level of $1-\alpha = 0.90$. As $n_{\text{split}}$ increases, the empirical coverage distribution becomes more concentrated around the target level, showing increased stability under repeated random splitting.  }    
    \label{fig: dist-coverage}
\end{figure}

In Fig.~\ref{fig: toy-cqr-mr} we show the resulting mass-radius curves computed using the polytropic EOS. As we can see, the prior curves cover a wide region in mass and radius, reflecting the weak constraints imposed by the priors. We then calculate the posterior mass-radius curves. As we show in Fig.~\ref{fig: toy-cqr-mr}, the likelihood penalizes soft EOSs that cannot support the observed NS masses. As a result, the posterior curves begin around $\approx 1.5\ M_{\odot}$, which is close to the lowest mass measurement used. However, there is no significant penalty or constraint on EOSs that support higher masses, and these remain allowed in the posterior. To quantify uncertainty in the mass-radius relation in a distribution-free and model-agnostic manner, we apply CQR as a postprocessing step to the posterior mass-radius curves. To construct the CQR bands, we consider the radius at each fixed mass and build prediction intervals using the posterior samples. Repeating this procedure across the full mass range yields the $90\%$ CQR band shown as the green shaded region in Fig.~\ref{fig: toy-cqr-mr}. As we can see, the CQR bands are narrower than the posterior because they are constructed to achieve a target coverage level (e.g., $90\%$). In particular, the CQR interval at each point is based on lower and upper quantiles, which exclude some posterior samples. Also, the CQR procedure identifies the most efficient region required to maintain the target coverage. This leads to bands that are narrower than the full posterior spread.

\begin{figure*}[t]
    \centering
    \includegraphics[width=0.32\textwidth]{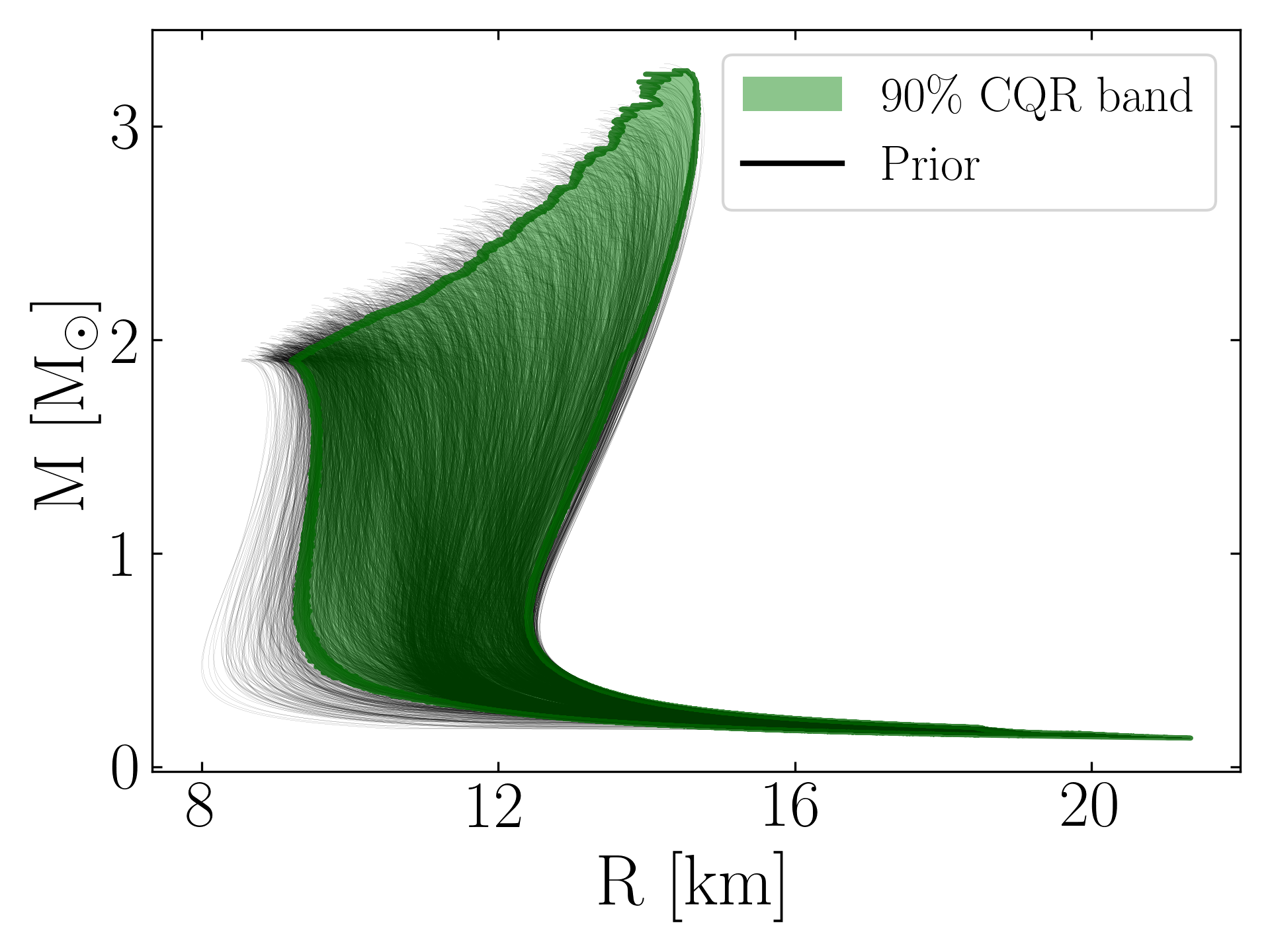}
    \includegraphics[width=0.32\textwidth]{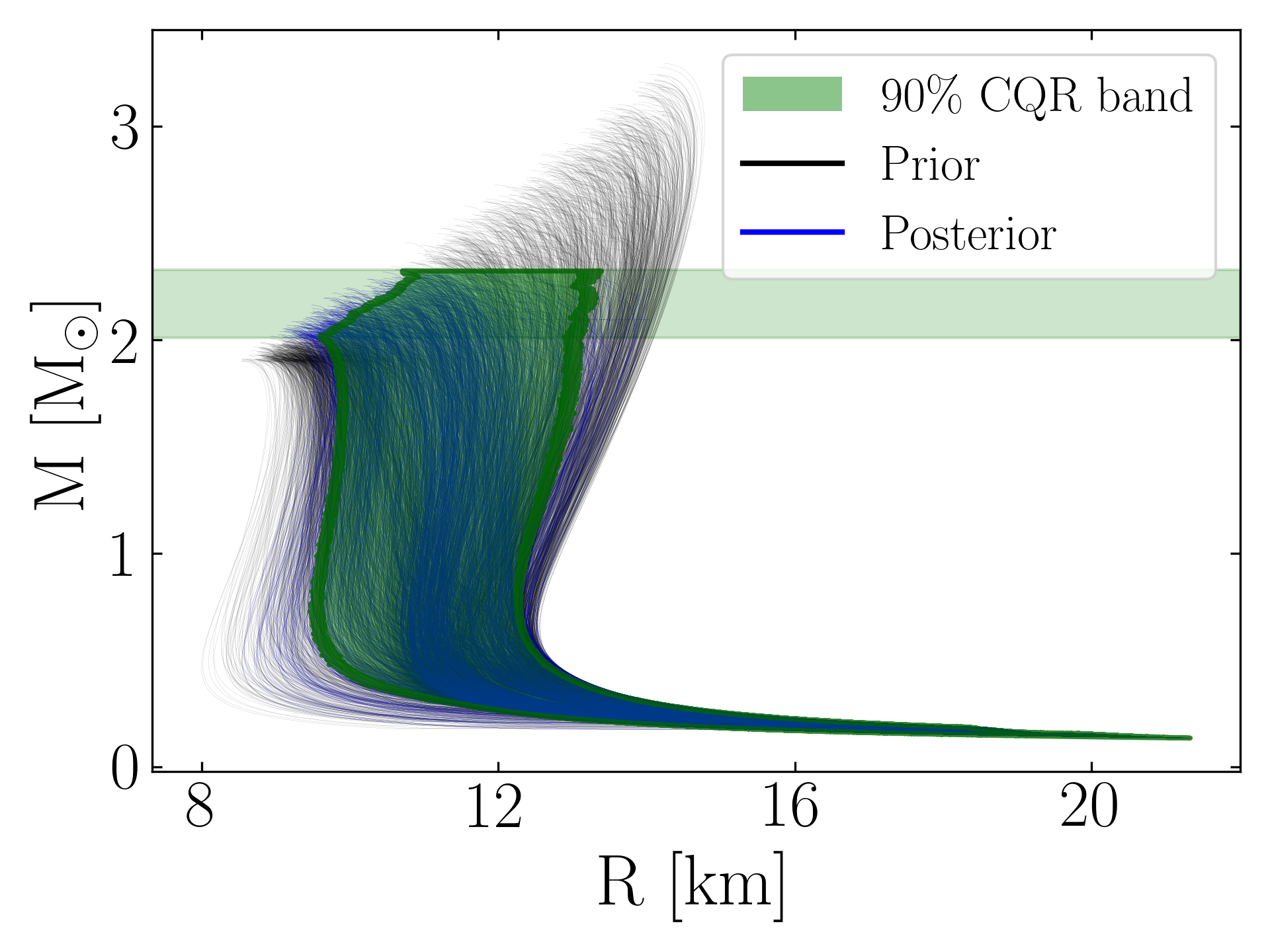}
    \includegraphics[width=0.32\textwidth]{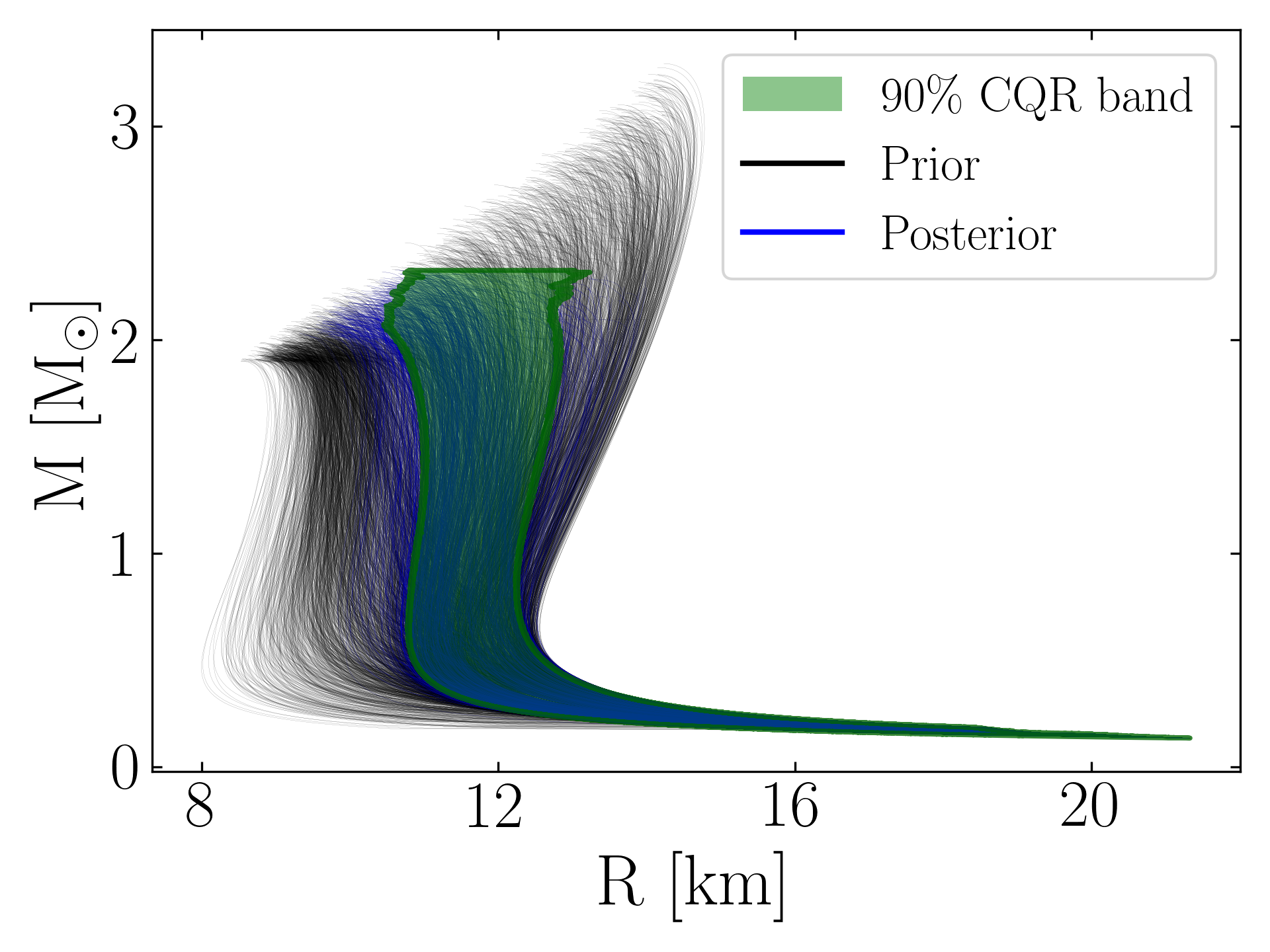}
    \caption{The 90\% CQR bands for the mass-radius relation of NSs. The black curves represent individual EOS samples provided by the NMMA collaboration \cite{dietrich_multi-messenger_2020}, and the green regions show the corresponding 90\% CQR prediction bands. In the left panel, we apply CQR to EOS samples from the NMMA analysis that are based on chiral EFT without additional filtering. In the middle panel, we apply CQR to the NMMA EOS samples that already satisfy the maximum mass constraint $M_{\text{max}} \leq 2.16^{+0.17}_{-0.15}$ at $2\sigma$ uncertainty, resulting in a narrower prediction band. In the right panel, we apply CQR to the NMMA EOS samples that incorporate the full set of astrophysical constraints included in the NMMA posterior, such as the maximum mass, the NICER observation of PSR J0030+0451 \cite{m_c_miller_et_al_psr_2019}, and gravitational wave information, including GW170817 \cite{b_p_abbott_et_al_gw170817_2017}, AT2017gfo \cite{b_abbott_et_al_multi-messenger_2017}, GRB170817A \cite{b_p_abbott_et_al_gravitational_2017}. The blue curves indicate EOSs that survive from the corresponding constraints.
     }
    \label{mr-intervals}
\end{figure*}
To assess the validity of the CQR prediction bands, we perform a model-checking study based on the empirical coverage. For a given error level $\alpha$, the full dataset at a fixed mass density of $2.0\times 10^{14} \text{g/cm$^3$}$ is randomly split into training, calibration, and validation datasets. For each split, the empirical coverage is evaluated on the validation data by checking whether the CQR prediction band contains the validation data points. We denote the number of validation data points by $n_{\text{valid}}$. It is important to note that, in this work, the empirical coverage is evaluated using validation data points drawn from the posterior distribution. Therefore, it refers to coverage with respect to the posterior sample-generating procedure and does not refer to the physical NS observable.
The number of times the dataset is randomly split into training, calibration, and validation sets is denoted by $n_{\text{split}}$.
This procedure is repeated over $n_{\text{split}}=1000$ independent random splits to obtain a distribution of empirical coverage values across a range of $1-\alpha$. In Fig.~\ref{fig: coverage}, the shaded regions represent the $68\% \ (1\sigma)$ and $95\% \ (1.96\sigma)$ confidence bands derived from the standard deviation across splits. The empirical coverage closely follows the ideal line and remains well within the confidence bands across the full range of $1-\alpha$, demonstrating that the CQR bands guarantee the desired coverage under minimal assumptions and highlighting the robustness and reliability of the CQR approach.

We also examine the conditional coverage of the CQR as a function of mass density using $10^4$ posterior samples. For comparison, we also include naive CP, which constructs prediction intervals using the absolute residual as the conformity score. A more detailed discussion of naive CP is provided in Ref.~\cite{dezdarani_conformal_2026}. In this analysis, the samples are split across the full range of mass density. We then evaluate the empirical coverage of the validation samples as a function of mass density and compare it with the target level ($1-\alpha= 0.90$).  As shown in Fig.~\ref{fig: conditional coverage}, naive CP overcovers at low densities and undercovers at high densities because it does not adapt to the variation in the spread of the posterior samples. In contrast, CQR adjusts the interval width as the variability changes and maintains empirical coverage close to the target level across the full density range.

In Fig.~\ref{fig: dist-coverage} we examine the stability of the empirical coverage distribution for CQR bands using the polytropic EOS at a fixed mass-density. We consider a dataset of $2000$ samples, which is randomly split into training, calibration, and validation sets $n_{\text{train}}=900$ for $n_{\text{split}}=100, 1000, 10000$. For each choice of $n_{\text{split}}$, the procedure is repeated over $n_{\text{trial}}=1000$ independent trials, where each trial uses freshly drawn samples. For every split, empirical coverage is evaluated by checking whether the CQR prediction band contains the validation data points, yielding a distribution of empirical coverage values rather than a single estimate.  As $n_{\text{split}}$ increases, the spread of the empirical coverage distribution narrows and becomes increasingly centered around the target coverage level $1- \alpha =0.90$. This behavior demonstrates that while conformal prediction guarantees marginal coverage for any finite calibration size, increasing the number of random splits leads to a more stable empirical coverage distribution. The empirical coverage being centered around the target level further supports the robustness and reliability of the CQR method.

\subsection{Postprocessing NMMA results}\label{sec: applications}
In this section, we demonstrate the application of the CQR method to the mass-radius relation of NSs using $\approx10^3$ EOS samples drawn from the posterior distribution published by the NMMA collaboration in Ref.~\cite{dietrich_multi-messenger_2020}. To create a smooth representation of the EOS samples, we interpolate the mass-radius data. For each EOS sample, we construct an interpolating function that returns the corresponding radius $R$ for any given mass $M$. This step allows us to evaluate the radius values at a common set of mass points across all EOSs, ensuring consistency with the toy-model analysis.

 \begin{figure*}[t]
    \centering
    \includegraphics[width=0.32\textwidth]{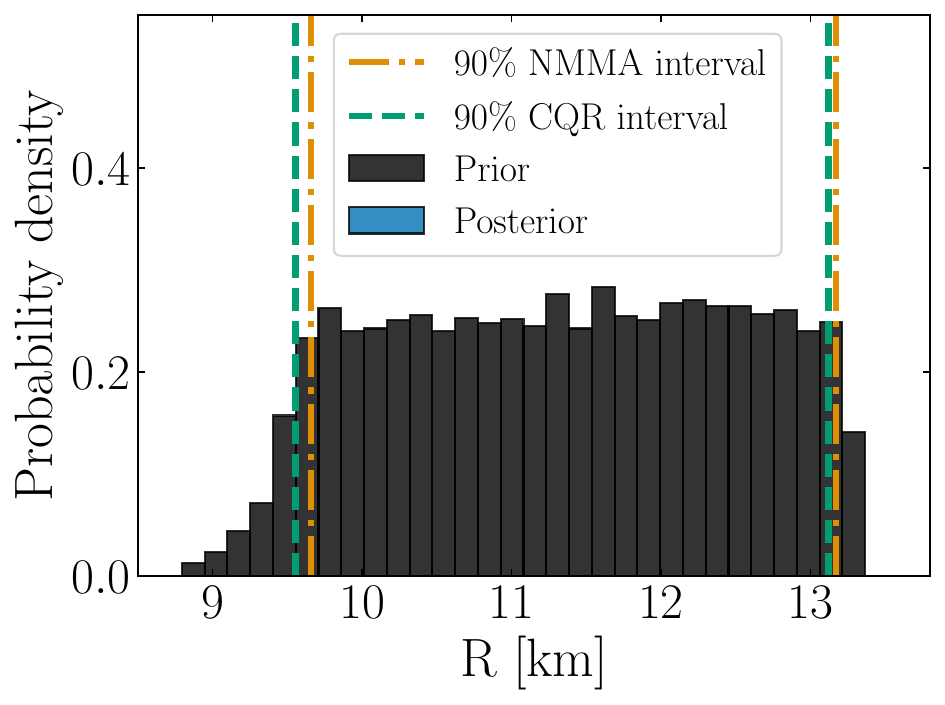}
    \hspace{0.01\textwidth}
    \includegraphics[width=0.32\textwidth]{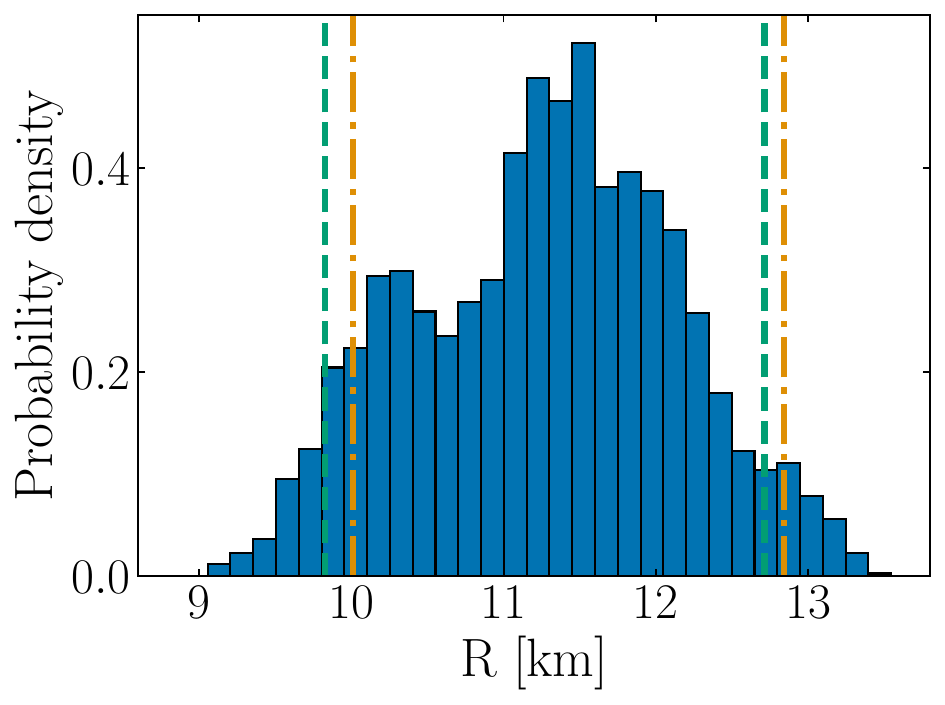}
    \includegraphics[width=0.32\textwidth]{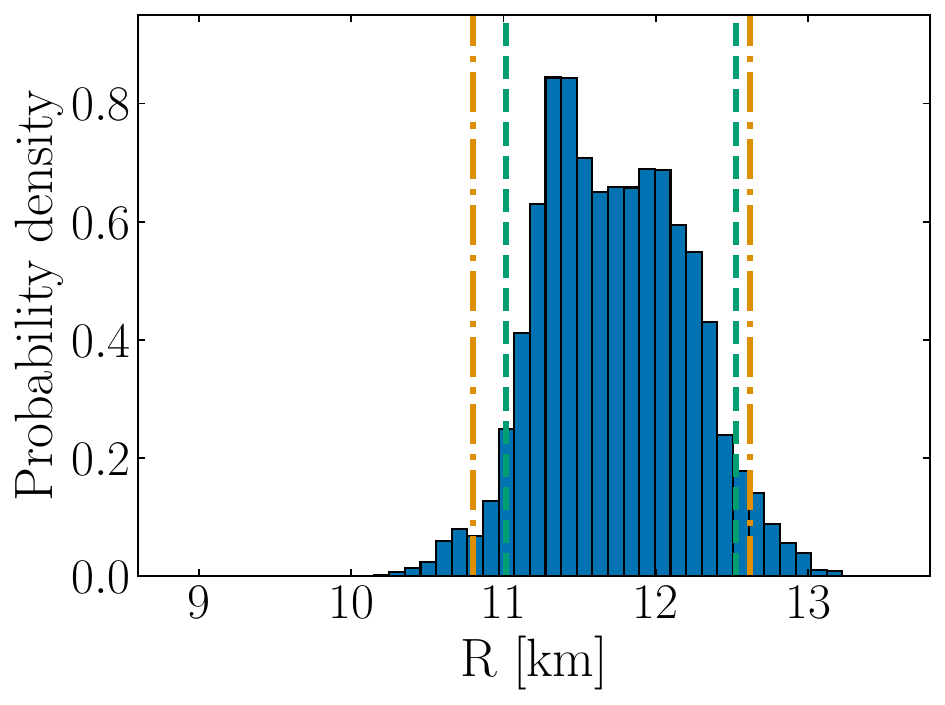}
    \caption{Distribution of NS radii at $1.4\ M_{\odot}$. The black left panel indicates the distribution calculated from EOSs derived from chiral EFT \cite{dietrich_multi-messenger_2020}, with a corresponding 90\% CQR interval of $11.35_{-1.80}^{+1.74}$ km (green dashed lines). The middle panel illustrates the distribution after imposing the maximum-mass constraint, resulting in a narrower 90\% CQR interval of $11.27_{-1.44}^{+1.40}$ km. The right panel corresponds to the fully constrained case, where the maximum mass requirement is combined with the NICER
     and gravitational wave event GW170817, together with electromagnetic counterparts  AT2017gfo and GRB170817A, yielding the narrowest 90\% CQR interval of $11.73_{-0.72}^{+0.80}$ km.}
    \label{r-hist}
\end{figure*}

After computing interpolated radius values for each EOS sample, we apply CQR to construct uncertainty bands for the radius as a function of mass.
At each fixed mass, we construct a CQR interval for the corresponding radius values. By repeating this process across the full mass range, we form the complete CQR band shown as the green bands in Fig.~\ref{mr-intervals}.  In the left panel of Fig.~\ref{mr-intervals}, we present the 90\% CQR band. The EOSs, shown as black curves, represent the low-density regime of nuclear matter up to $1.5n_{\text{sat}}$, with $n_{\text{sat}} = 0.16$ fm$^{-3}$ is the nuclear saturation density. 
In the middle panel of Fig.~\ref{mr-intervals}, we consider EOS samples drawn from the posterior distribution that already satisfy the maximum-mass constraint provided in the NMMA analysis. The NMMA analysis combines radio pulsar observations with gravitational wave constraints to infer an upper and lower bound on the maximum mass of NSs, $M_{\text{max}}\leq 2.16^{+0.17}_{-0.15}$ at the $2\sigma$ uncertainty level \cite{dietrich_multi-messenger_2020}. In that work, the different constraints are incorporated using a likelihood based on the error function. We then apply CQR to these restricted samples, which leads to narrower prediction bands compared to the previous case, reflecting the reduced spread of the underlying EOS samples.
\begin{figure*}
    \centering
    \includegraphics[width=0.32\textwidth]{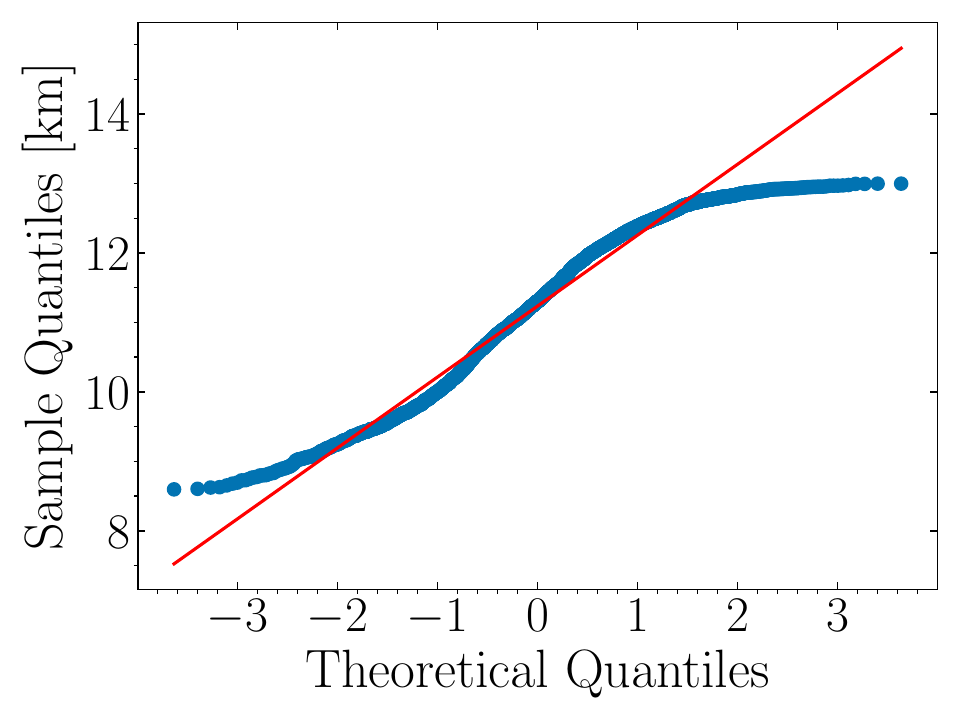}
    \hspace{0.01\textwidth}
    \includegraphics[width=0.32\textwidth]{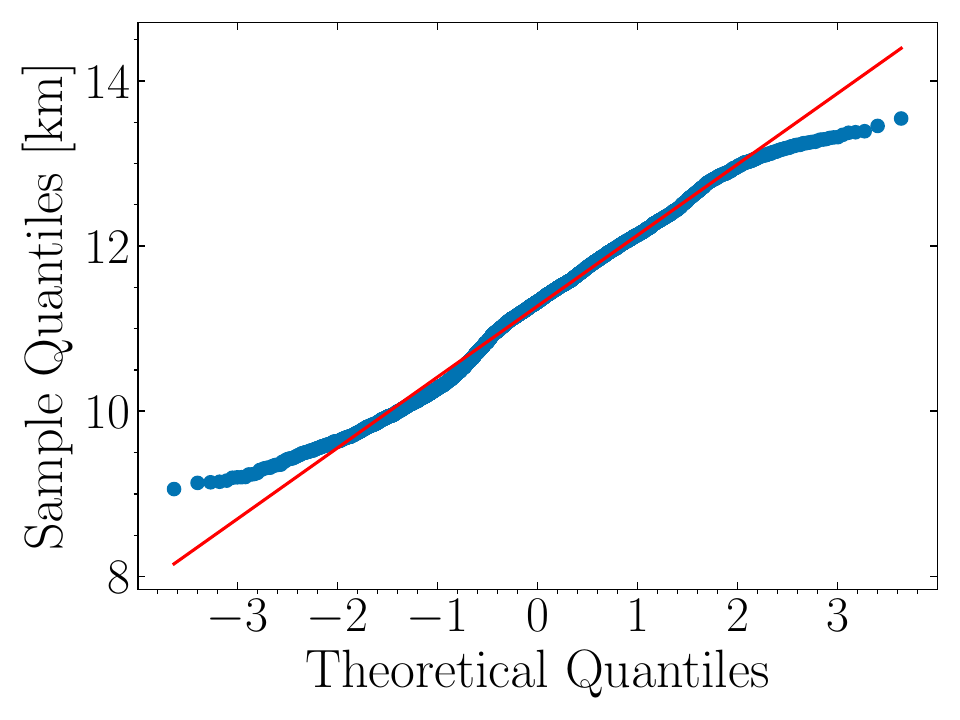}
    \includegraphics[width=0.32\textwidth]{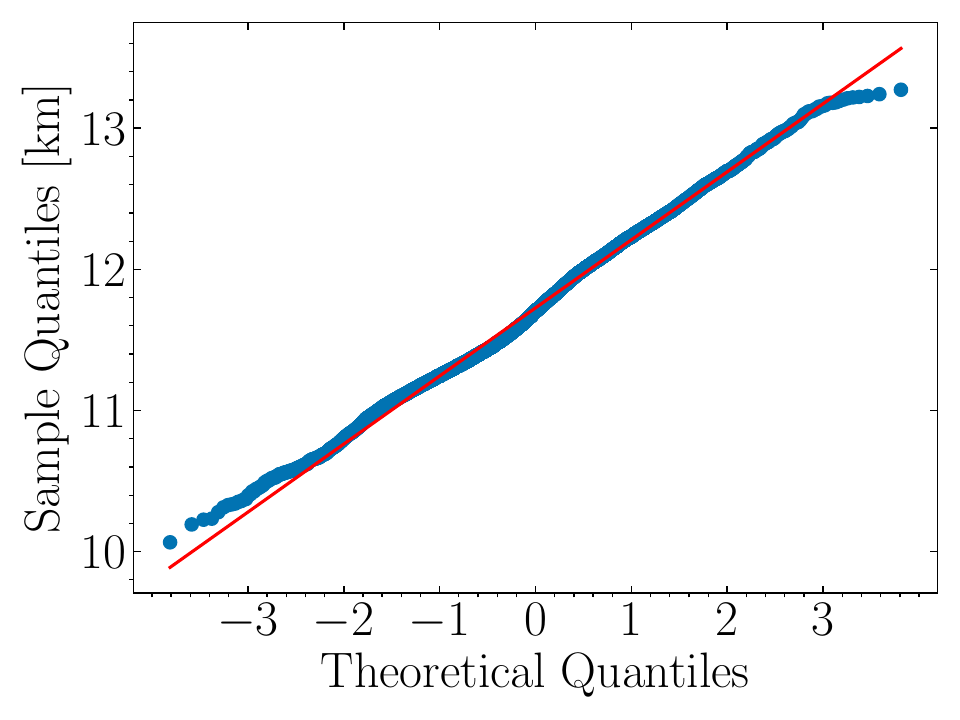}
    \caption{Q-Q plots of the radius samples at a fixed mass of $M = 1.4\ M_{\odot}$ for the three EOS selections shown in Fig.~\ref{mr-intervals}. The horizontal axis indicates the theoretical quantiles, while the vertical axis demonstrates the empirical quantiles of the radius samples. The red diagonal line indicates perfect agreement with a normal distribution. Deviations from this line reveal non-Gaussian features in the radius distributions, motivating the use of distribution-free methods such as CQR for uncertainty quantification.}
    \label{fig: qq-rm}
\end{figure*} 
Finally, in the right panel of Fig.~\ref{mr-intervals}, we consider the full set of astrophysical constraints provided in the NMMA analysis, including the mass-radius posterior based on NICER observations
of PSR J0030+0451 \cite{m_c_miller_et_al_psr_2019}, gravitational-wave events, such as  GW170817 \cite{b_p_abbott_et_al_gw170817_2017}, and its electromagnetic counter parts AT2017gfo \cite{b_abbott_et_al_multi-messenger_2017}, and GRB170817A \cite{b_p_abbott_et_al_gravitational_2017}. We apply CQR to the EOS samples that already satisfy these constraints, and use surviving EOSs to construct the final $90\%$ CQR prediction band, which results in a narrower interval than in the first two panels.

In Fig.~\ref{mr-intervals}, the 90\% CQR bands are seen to capture the majority of EOS curves. To better assess the performance of the CQR method, we focus on the commonly studied mass of  $1.4 \ M_{\odot}$. In Fig.~\ref{r-hist} we present histograms of the radii at $M = 1.4 M_{\odot}$ for each of the three cases considered in Fig.~\ref{mr-intervals}.
In the left panel, the radius distribution calculated from the NMMA EOS samples based on chiral EFT EOSs without astrophysical filtering is broad and covers a wide range of radii, yielding a 90\% CQR interval of $11.35_{-1.80}^{+1.74}$ km. In the middle panel, conditioning on the maximum mass constraint removes EOSs that cannot support the required mass, resulting in a narrower radius distribution and a reduced 90\% CQR interval of $11.27_{-1.44}^{+1.40}$ km. Finally, in the right panel, incorporating the full set of astrophysical constraints included in the NMMA posterior yields the narrowest 90\% CQR interval of $11.73_{-0.72}^{+0.80}$ km, indicating how astrophysical filtering steps tighten the CQR interval. For comparison, the NMMA collaboration reports an NS radius at $M = 1.4 \ M_\odot$ of $11.67_{-0.87}^{+0.95}$ km at the $90\%$ credibility level. The final CQR interval that we calculated here is smaller than the NMMA interval.
These results demonstrate how the CQR method was constructed from the posterior samples. The posterior already encodes the probability density of EOSs. The CQR band is calculated by imposing a target coverage level on posterior samples. 

\begin{figure}[b]
    \centering
    \includegraphics[width=0.5\textwidth]{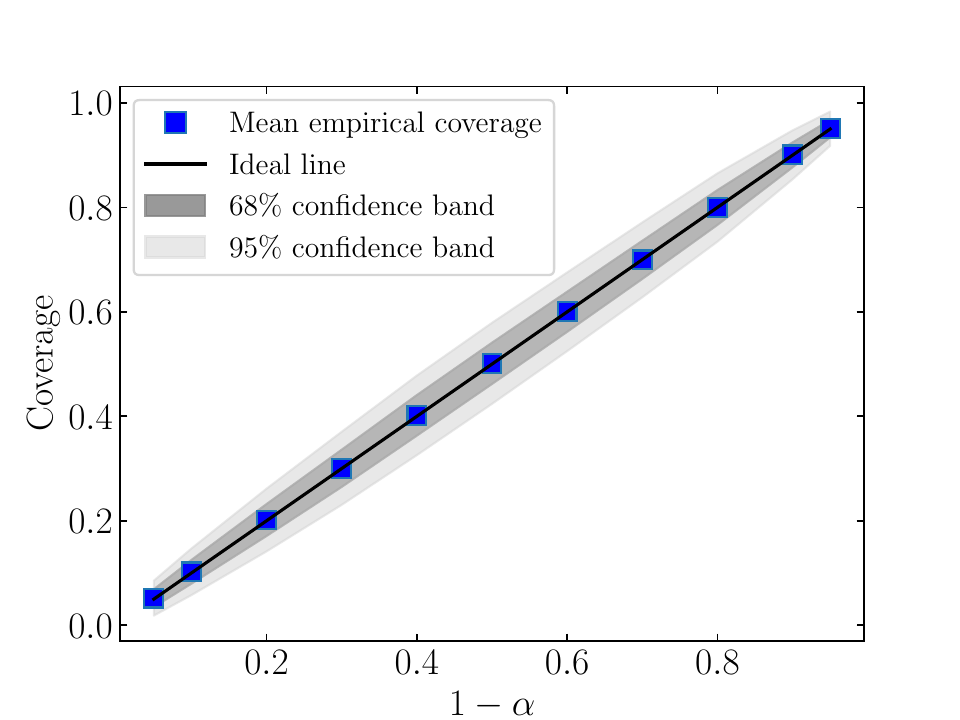}
    \caption{Empirical coverage of CQR prediction intervals for radius samples at a fixed mass of $1.4\ M_{\odot}$, using EOSs that satisfy the full astrophysical constraints. For each target level of $1 - \alpha$, the mean empirical coverage is evaluated over $n_{\text{split}}=1000$ independent random splits. The mean empirical coverage closely follows the ideal coverage line across a broad range of $1-\alpha$. This behavior reflects the distribution-free coverage guarantees of CP, which remain valid for realistic EOSs, including cases where the underlying radius distributions exhibit skewness or heavy tails.}    
    \label{fig: coverage-r-nmma}
\end{figure} 

To further assess the statistical behavior of the radius distributions, we examine the quantile-quantile (Q-Q) plots of the radius samples at a fixed mass of $1.4\ M_{\odot}$, as shown in Fig.~\ref{fig: qq-rm}. The Q-Q plot provides a diagnostic tool to evaluate how closely the empirical distribution of the samples follows a normal distribution by comparing their empirical quantiles with the corresponding theoretical quantiles. In Fig.~\ref{fig: qq-rm}, each panel corresponds to one of the three cases discussed in Fig.~\ref{mr-intervals}: the chiral EFT-based EOS set, the subset satisfying the maximum-mass constraint, and the fully constrained EOS set incorporating all astrophysical information. If the sample distribution were exactly normal, the empirical quantiles would lie along the diagonal reference line. Deviations from this line, therefore, indicate departures from normality, such as skewness or heavy tails. Across all three cases, the Q-Q plots illustrate noticeable deviations from the diagonal line. This reflects the non-Gaussian behavior of the radius distributions. The CQR method provides prediction intervals with guaranteed coverage without relying on a specific distributional form, making CQR suitable for constructing uncertainty bands for this setting. The Q-Q analysis thus complements the CQR results by illustrating why a distribution-free approach is appropriate for constructing an uncertainty band with coverage guarantees in the mass-radius relation.

To validate the CQR prediction intervals in the fully constrained astrophysical setting, we repeat the same empirical coverage analysis introduced in Fig.~\ref{fig: coverage} for the polytropic EOS. The model-checking procedure, including the data splitting strategy, is the same as that described in the toy model section. In Fig.~\ref{fig: coverage-r-nmma} we demonstrate that the mean empirical coverage of the radius samples closely follows the ideal coverage line across the full range of $1- \alpha$. This indicates that the coverage guarantees observed in the toy model analysis continue to hold when the method is applied to realistic NS EOSs constrained by astrophysical observations. Importantly, these results are obtained without relying on any specific distribution, i.e., regardless of the fact that the fully constrained radius distributions exhibit clear deviations from normality.

\subsection{Quantum Monte Carlo}

In this application, we consider the use of the CQR method for the EOS of pure neutron matter (PNM) calculated with Quantum Monte Carlo (QMC). The QMC method is a powerful, stochastic approach to finding the ground-state energy of a system. It achieves this by applying an imaginary-time propagator to a physically informed trial wavefunction, which projects out the ground state of the system~\cite{carlson_quantum_2015}. In particular, we consider results calculated using auxiliary-field diffusion Monte Carlo (AFDMC), which enables accurate calculations of the energy per particle of PNM with realistic nuclear interactions \cite{Gandolfi:2009fj, Gezerlis:2014zia, Tews:2015ufa, Lynn:2015jua, Tews:2018kmu, Lonardoni:2019ypg, Curry:2024gcz, Tews:2024owl}. 

The calculations considered here, published in Ref.~\cite{armstrong_emulators_2025}, are based on $\approx10^5$ energy per particle samples at fixed density, obtained from emulator predictions trained on AFDMC results. These samples encode the combined theoretical and statistical uncertainties associated with the underlying nuclear interactions and many-body calculations.

\begin{figure}[t]
    \centering
    \includegraphics[width=0.5\textwidth]{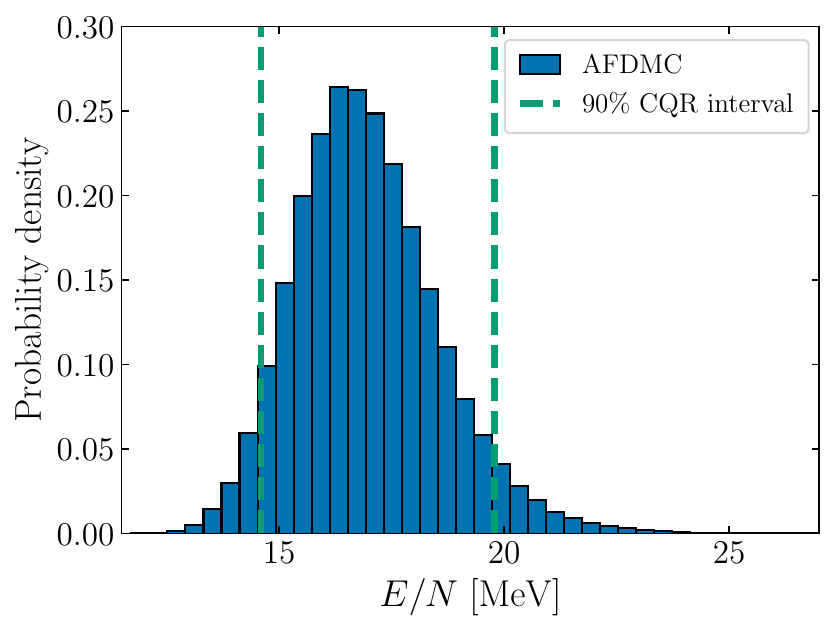}
   
    \caption{Distribution of the neutron matter energy at a density of $n=0.16 \ \text{fm}^{-3}$. The distribution is constructed from $\approx10^5$ samples calculated from emulator predictions trained on AFDMC calculations \cite{armstrong_emulators_2025}. The green dashed lines indicate the corresponding $90\%$ CQR intervals.}
    \label{fig: density-hist}
\end{figure}

\begin{figure}[b]
    \centering
    \includegraphics[width=0.5\textwidth]{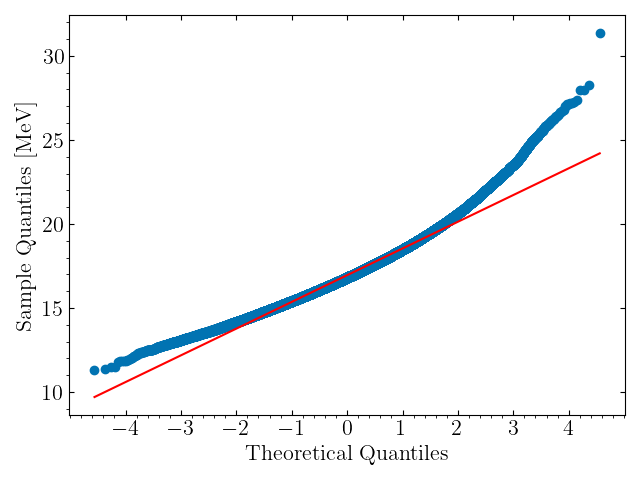}

    \caption{Q-Q plots of the neutron matter energy per particle at a density of $n=0.16 \  \text{fm}^{-3}$. The empirical quantiles of the energy samples are compared to the theoretical quantiles of a normal distribution. Observed deviations from this line reveal non-Gaussian features in the underlying distributions.  }
    \label{fig: qq-density}
\end{figure}

In Fig.~\ref{fig: density-hist} we show the distribution of the
neutron-matter energy per particle at a density of $n=0.16 \ \text{fm}^{-3}$, which demonstrates deviations from Gaussian distributions, including extended tails that become more significant at higher densities. The corresponding Q-Q plot in Fig.~\ref{fig: qq-density} provides a complementary diagnostic by comparing the empirical quantiles of the samples to those of a normal distribution. Deviations from the diagonal reference line are observed, particularly in the tails, confirming that the underlying distribution is not well described by Gaussian behavior. These results motivate us to use CQR as a postprocessing step to construct prediction intervals.

\begin{figure}
    \includegraphics[width=0.5\textwidth]{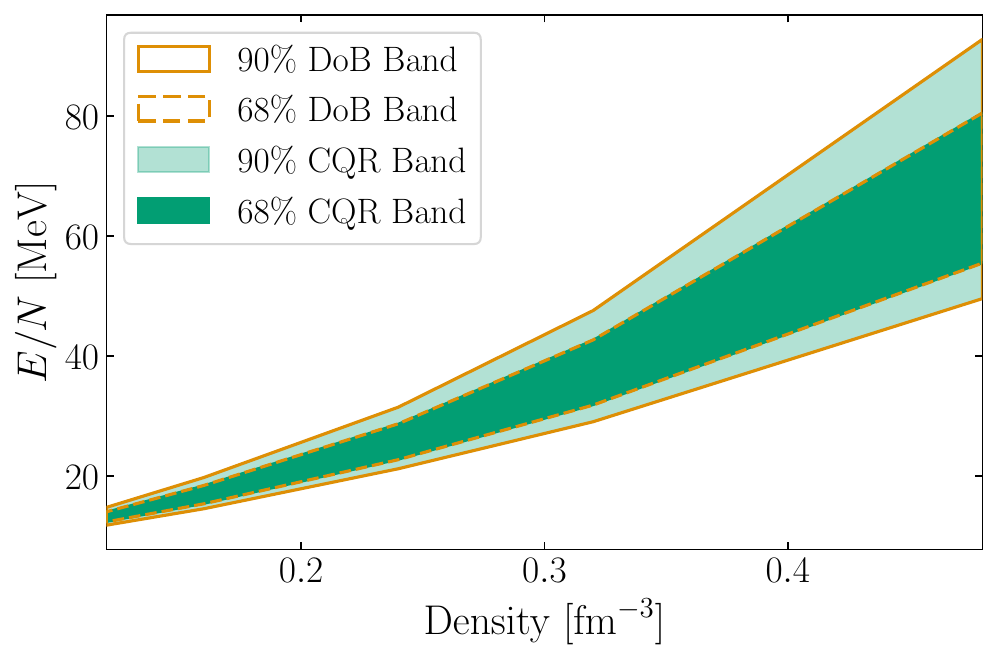}
    \caption{A comparison between the 68\% and 90\% CQR prediction intervals and the corresponding DoB uncertainty bands for the neutron matter energy per particle as a function of number density. The DoB bands are calculated from the emulator-based Bayesian framework \cite{armstrong_emulators_2025}, while the CQR intervals are constructed as a postprocessing step from the energy samples. For the large sample size used here, approximately $10^5$, the CQR and DoB intervals are nearly identical across the density range.}
    \label{fig: cqr-dob-eos}
\end{figure}
\begin{figure}[b]
    \includegraphics[width=0.45\textwidth]{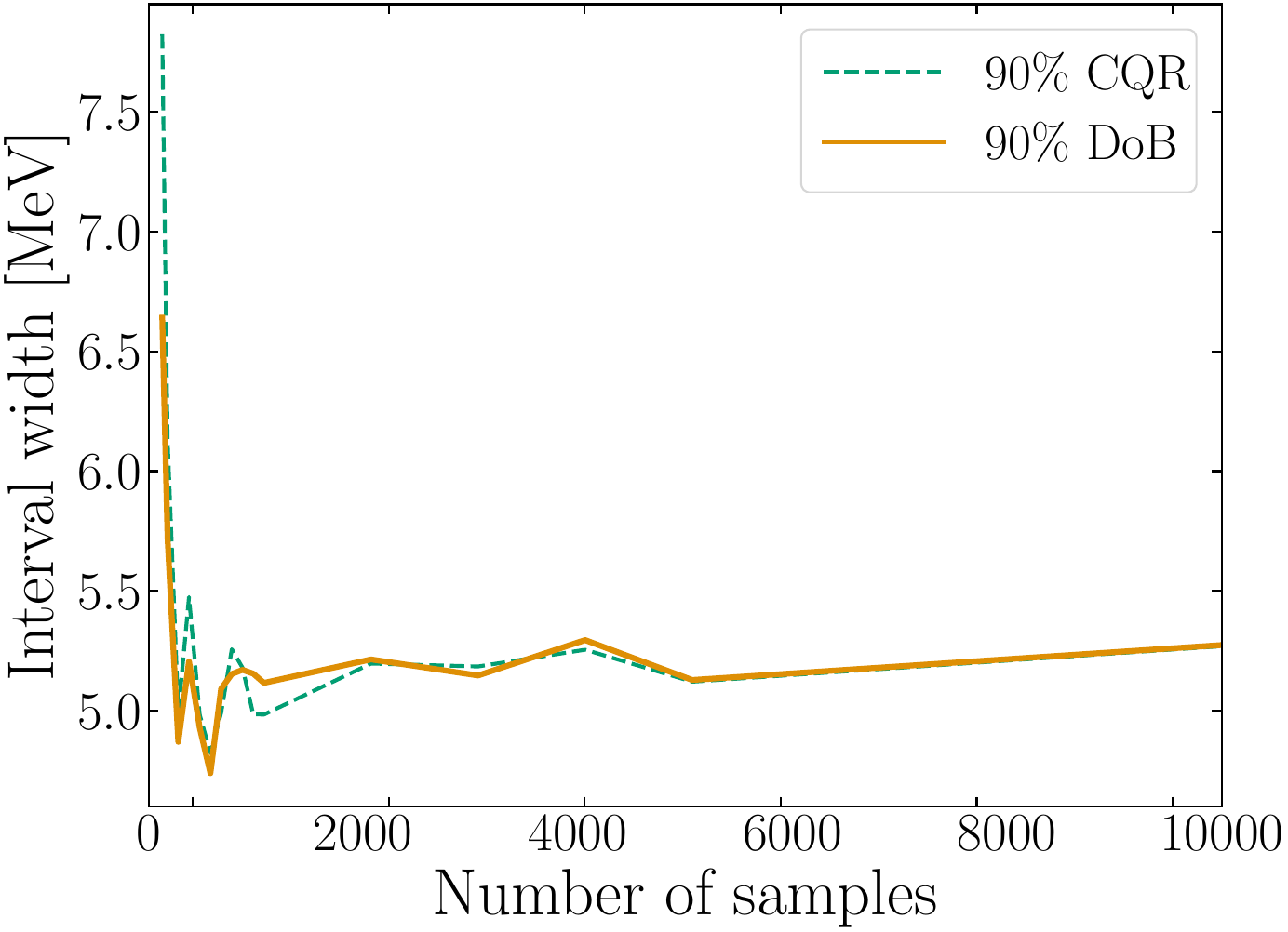}
  
    \caption{Interval width of the CQR and DoB bands for the neutron matter energy per particle at a fixed density of $0.16\ \text{fm}^{-3}$, as a function of the number of samples. At small sample sizes, both intervals fluctuate and can differ from each other.  As the number of samples increases, both intervals converge to the same value, consistent with the close agreement observed in Fig.~\ref{fig: cqr-dob-eos}  at $\approx 10^5$ samples. This comparison indicates how many samples are required for DoB intervals to match the coverage-guaranteed CQR results.}
    \label{fig: qmc-interval_width}
\end{figure}

\begin{figure}[b]
    \centering
    \includegraphics[width=0.45\textwidth]{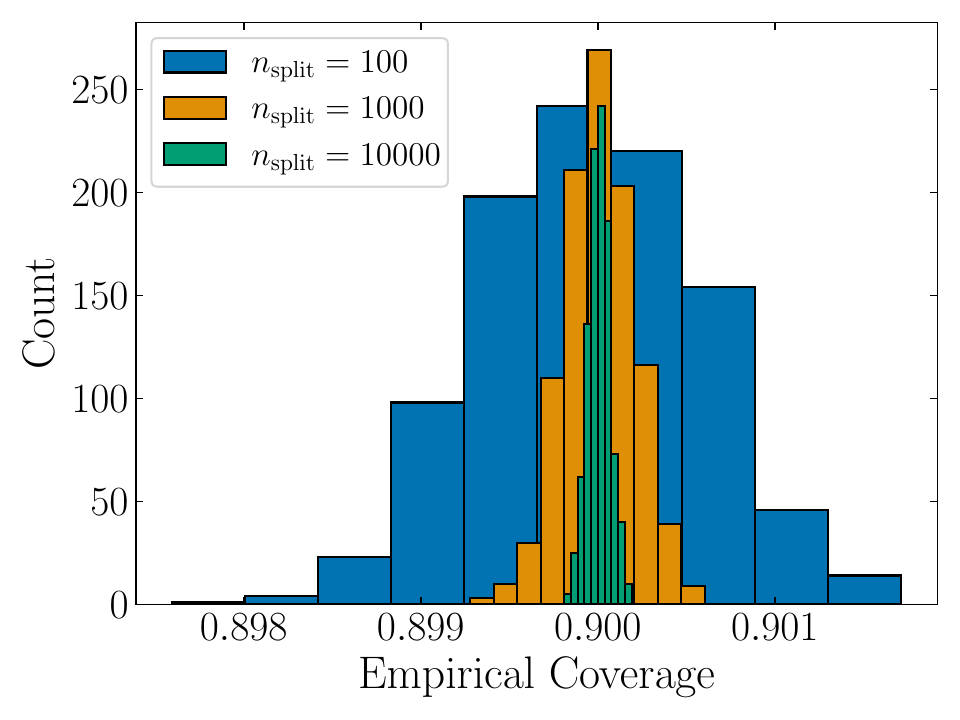}
    \hspace{0.05\textwidth}
    \includegraphics[width=0.45\textwidth]{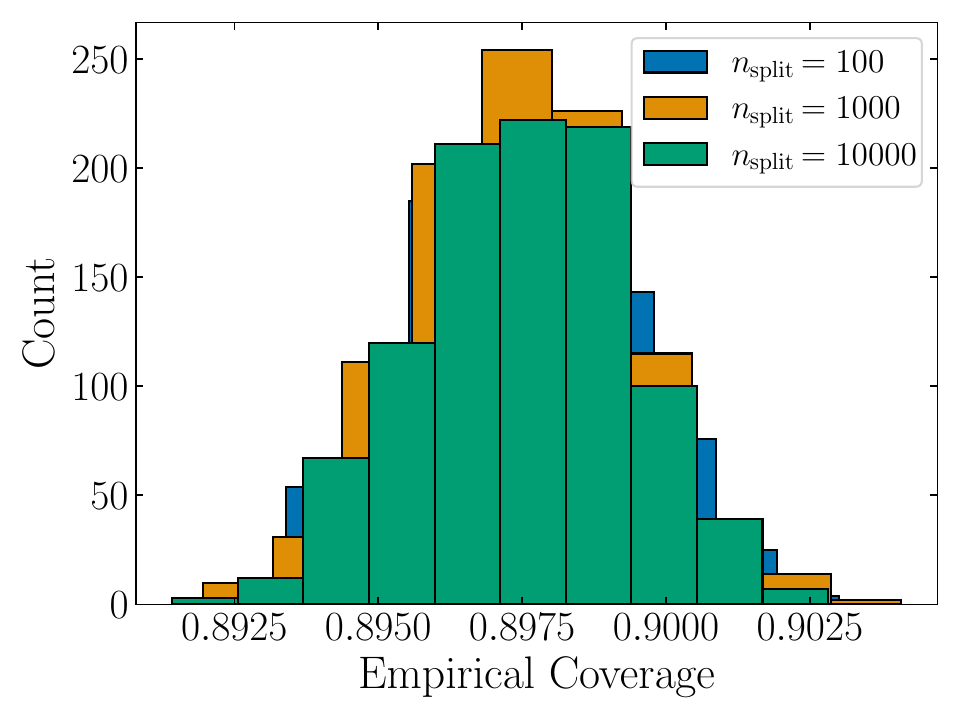}
    \caption{Empirical coverage distribution for the neutron matter energy per particle at a fixed density of $n=0.16$ fm$^{-3}$ for three different values of  ($n_{\text{split}}=100,1000,10000$). The dataset contains $25000$ energy samples and is randomly split into $ n_{\text{calib}}=11000$,  $n_{\text{train}}=11000$, and $n_{\text{valid}}=3000$. The upper panel represents the CQR results, and the lower panel indicates the corresponding DoB results. As  $n_{\text{split}}$ increases, the CQR empirical coverage distribution becomes narrower and centered around the target coverage $0.90$. In contrast, the DoB interval is fixed, so the corresponding distributions do not narrow in the same way and remain slightly shifted away from the target coverage.}
    \label{fig: qmc-coverage-DoB,CQR}
\end{figure}

\begin{figure}[b]
    \includegraphics[width=0.5\textwidth]{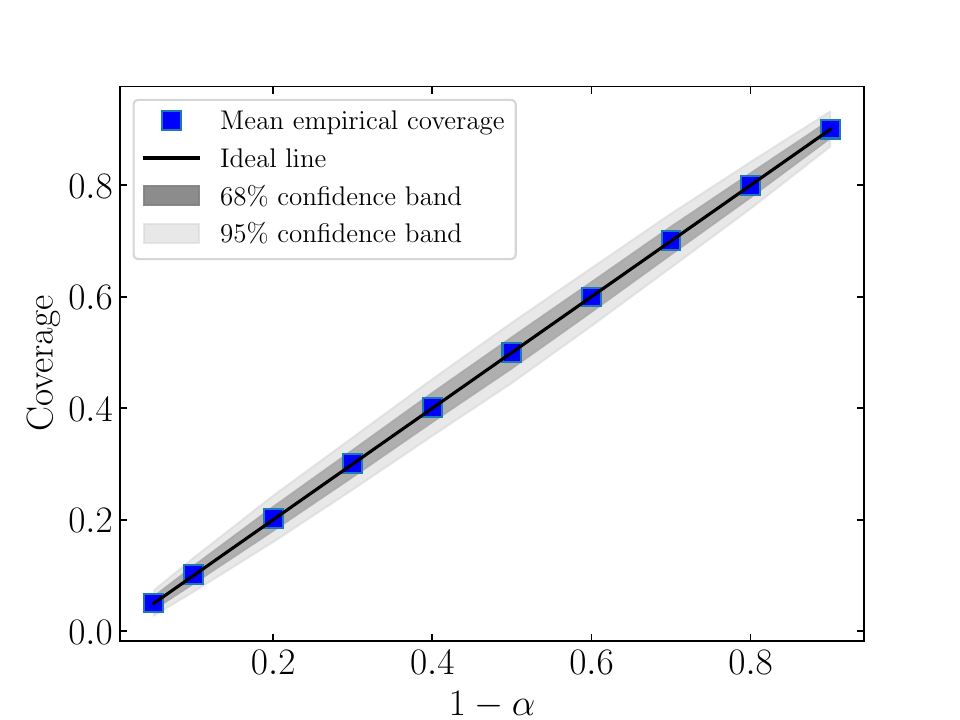}
    \caption{Empirical coverage of CQR prediction intervals for neutron matter energy per particle at a fixed number density $n=0.16\ \text{fm}^{-3}$ as a function of the target coverage level $1-\alpha$. For each value of $1-\alpha$, the squares indicate the mean empirical coverage calculated using $n_{\text{split}}=1000$. In this case, as in the toy model and the previous application, the mean empirical coverage closely follows the ideal coverage line, confirming that CQR achieves its guaranteed coverage. The shaded bands of empirical coverage around the ideal line reflect the stability of the method for this setting.}    
    \label{fig: coverage-density16}
\end{figure} 

In Fig.~\ref{fig: cqr-dob-eos} we compare the CQR prediction intervals with the corresponding degree of belief (DoB) bands for the neutron matter energy per particle as a function of density. The DoB bands shown here are calculated from the emulator-based Bayesian framework \cite{armstrong_emulators_2025}, while the CQR intervals are constructed as a postprocessing step using the energy samples. With approximately $10^5$ samples used in this analysis, the CQR intervals closely match the corresponding DoB bands across the full density range. This agreement indicates that the underlying energy distributions are well sampled. However, the two intervals have different statistical interpretations: the DoB bands have a Bayesian probabilistic interpretation, whereas the CQR bands have a frequentist interpretation and are calibrated to achieve the target coverage for future data. This distinction becomes more important in the small sample regime, where the CQR coverage guarantee remains valid without assuming a specific distributional form.

To further investigate the relationship between the two methods, we examine how the interval width evolves as a function of sample size, as shown in Fig.~\ref{fig: qmc-interval_width}. At small sample sizes, both CQR and DoB fluctuate considerably and can differ from one another. However, as more samples are included, the two intervals converge to the same value. This behavior is robust across independent runs and confirms that in the large-sample regime, the CQR interval width stabilizes and closely matches the corresponding DoB interval width.

In Fig.~\ref{fig: qmc-coverage-DoB,CQR}, we compare the empirical coverage distributions calculated from CQR and DoB intervals for the neutron matter energy per particle at fixed density $n=0.16$ fm$^{-3}$. This analysis is based on a dataset of $25000$ energy samples, which is randomly divided into $ n_{\text{calib}}=11000$,  $n_{\text{train}}=11000$, and $n_{\text{valid}}=3000$. For the CQR case, shown in the top panel, as $n_{\text{split}}$ increases, the distribution becomes narrower and more concentrated around the target coverage of $0.90$. For the DoB intervals, shown in the bottom panel, the three distributions overlap rather than narrowing since the DoB interval is fixed. Furthermore, the DoB distributions are not centered around the target coverage of $0.90$, representing a slight offset, in contrast to the CQR distributions, which remain well centered around the target level.

In Fig.~\ref{fig: coverage-density16} we present the empirical coverage of CQR intervals for the neutron matter energy per particle at a fixed density $n=0.16\ \text{fm}^{-3}$. As in the toy model and the previous application considered in this work, the empirical coverage closely follows the ideal empirical coverage across all target levels, confirming that the CQR achieves its guaranteed coverage. 

\section{Summary and outlook}

In this work, we have demonstrated how CP, and in particular CQR, can be integrated with Bayesian inference to provide reliable and distribution-free uncertainty quantification for the EOS of NSs and related observables. By treating CQR as a postprocessing step applied to posterior samples, this approach preserves the Bayesian philosophy while providing finite-sample coverage guarantees that do not rely on assumptions about the underlying distributions. Using a Bayesian model based on a polytropic EOS and the TOV equations, we have performed Bayesian inference to calculate posterior distributions of the EOS parameters and the corresponding NS observables. Based on these posterior samples, we construct $90\%$ CQR prediction bands for both the EOS and the mass-radius relation and validate the CQR framework through model-checking studies. The empirical coverage closely follows the ideal line across a range of target coverage. Considering different random splits demonstrates the stability of the resulting prediction bands, emphasizing the robustness of CQR.

We then applied the method to realistic mass-radius relations of NSs derived from chiral EFT provided by the NMMA collaboration. By incorporating observational information, the resulting $90\%$ CQR bands become narrower. For the commonly used mass  $1.4\ M_{\odot}$, the final CQR prediction interval for the NS radius is $11.73_{-0.72}^{+0.80}$ km after applying the full set of astrophysical constraints, which is comparable with the NMMA results of $11.67_{-0.87}^{+0.95}$. We have also shown that the CQR intervals adapt to non-Gaussian distributions of the NS radius, as supported by QQ plots. Repeating the empirical coverage analysis in this realistic setting confirms that the coverage behavior observed in the toy model remains consistent for chiral EOSs constrained by observational data.

Finally, we applied CQR to the EOS of PNM obtained from QMC calculations. The resulting prediction intervals capture the asymmetric and heavy-tailed features of the energy distributions and demonstrate empirical coverage consistent with the target coverage. Comparison with the Bayesian DoB intervals illustrates the complementary role of CQR as a distribution-free method that adapts to the shape of the data.

Overall, these results show that CQR provides a distribution-free and model-agnostic method for constructing uncertainty bands in NS physics. Without relying on assumptions about the underlying distribution, it can be combined with Bayesian inference. 
The method applies to EOS inference and other astrophysical observables and offers a practical approach to uncertainty quantification with guaranteed coverage.

\section*{Acknowledgments}
The authors would like to thank Ingo Tews for a careful reading of the manuscript and providing detailed feedback, as well as the NMMA collaboration for making their data publicly
available.  H.Y.D., R.C., and A.G. were supported by the Natural Sciences and Engineering Research Council (NSERC) of Canada and the Canada Foundation for Innovation (CFI). C.L.A. was supported by the Michigan State University Distinguished Fellowship (UDF) from the Michigan State University Graduate School. Computational resources have been provided by Compute Ontario through the Digital Research Alliance of Canada, and by the National Energy Research Scientific Computing Center (NERSC), which is supported by the U.S. Department of Energy, Office of Science, under contract No. DE-AC02-05CH11231.

\bibliography{bib}
\end{document}